# Visualization of Software and Systems as Support Mechanism for Integrated Software Project Control

Peter Liggesmeyer[1,2], Jens Heidrich[1], Jürgen Münch[1], Robert Kalcklösch[2], Henning Barthel[1], Dirk Zeckzer[2]

[1]Fraunhofer IESE, Fraunhofer Platz 1, 67663 Kaiserslautern, Germany
{peter.liggesmeyer, jens.heidrich, juergen.muench, henning.barthel}@iese.fraunhofer.de

[2]TU Kaiserslautern, Post Office Box 3049, 67653 Kaiserslautern, Germany
{kalckloesch, zeckzer}@informatik.uni-kl.de

**Abstract.** Many software development organizations still lack support for obtaining intellectual control over their software development processes and for determining the performance of their processes and the quality of the produced products. Systematic support for detecting and reacting to critical process and product states in order to achieve planned goals is usually missing. One means to institutionalize measurement on the basis of explicit models is the development and establishment of a so-called Software Project Control Center (SPCC) for systematic quality assurance and management support. An SPCC is comparable to a control room, which is a well known term in the mechanical production domain. One crucial task of an SPCC is the systematic visualization of measurement data in order to provide context-, purpose-, and role-oriented information for all stakeholders (e.g., project managers, quality assurance managers, developers) during the execution of a software development project. The article will present an overview of SPCC concepts, a concrete instantiation that supports goal-oriented data visualization, as well as examples and experiences from practical applications.

## 1 Introduction

The complexity of software-intensive systems and development projects continues to increase. One major reason is the ever-increasing complexity of functional as well as non-functional software and systems requirements (e.g., reliability or time constraints for safety critical systems). The more complex the requirements, the more people are usually involved in meeting them, which further increases the complexity of controlling and coordinating a project. This, in turn, makes it even harder to develop the system according to plan (i.e., matching time, quality, and budget constraints). Project control issues are very hard to handle. Many software development organizations still lack support for obtaining intellectual control over their software development pro-jects and for determining the performance of their processes and the quality of the produced products. Systematic support for detecting and reacting to critical process and product states in order to achieve planned goals and quality is usually missing [15]. One way to support effective control of software development projects is the use of basic engineering principles [7], [19], with particular attention to the monitoring and analysis of actual product and process states, the comparison of actual states with planned states, and the initiation of any necessary corrective actions during project execution. Effectively applying these principles requires the collection, interpretation, and appropriate visualization of measurement data according to a previously measurement goals and plans in order to provide stakeholders with up-to-date information about the project state. One major challenge is to adequately (and partially integrated) visualize process and product properties during project execution so that informed decisions can be made by the relevant stakeholders (such as project managers, quality assurance personnel). This addresses, for instance, early warning mechanisms that recognize insufficient quality characteristics of development products or the ability to generate accurate effort and cost predictions.

In the aeronautical domain, air traffic control systems are used to ensure the safe operation of commercial and private aircraft. Air traffic controllers use these systems to coordinate the safe and efficient movement of air traffic (e.g., to make certain that planes stay a safe distance apart or to minimize delays). These systems collect and visualize all critical data (e.g., the distance between two planes, the planned arrival and departure times) in order to support decisions by air traffic controllers. Software project control requires an analogous approach that is tailored to the specifics of the process being used (e.g., its non-deterministic, concurrent, and distributed nature). A Software Project Control Center (SPCC) [15] is a control system for software development that collects all data relevant to project control, interprets and analyzes the data according to the project's control needs, visualizes the data for different project roles, and suggests corrective actions in the case of plan deviations. An SPCC could also support the packaging of data (e.g., as predictive models) for future use and contribute to an improvement cycle spanning a series of projects. Controlling a project means ensuring the satisfaction of project objectives by monitoring and measuring progress regularly in order to identify variances from the plan during project execution, so that corrective action can be taken when necessary [17]. Planning is the basis for project control and defines expectations, which can be checked during project execution. Project control is driven by different role-oriented needs. We define control needs as a set of role-dependent requirements for obtaining project control. A project manager needs different kinds of data, data of different granularity, or different data visualizations than a quality assurance manager.

In this article, we want to illustrate selected existing project control approaches (Section 2), and then focus on a concrete instantiation that supports goal-oriented data visualization, the so called Specula approach (Section 3). Afterwards, we will present advanced visualization mechanisms used for controlling risks and quality of development projects and selected lessons learned from their application (Section 4). Finally, we will give a summary and illustrate future research fields (Section 5).

## 2   Related Work

An overview of the state of the art in Software Project Control Centers can be found in [15]. Most of the existing, rather generic, approaches for control centers offer only partial solutions. Especially purpose- and role-oriented usages based on a flexible set of techniques and methods are not comprehensively supported. In practice, many companies develop their own dashboards (mainly based on Spreadsheet applications) or use dashboard solutions that provide a fixed set of predefined functions for project control (e.g., deal with product quality only or solely focus on project costs) and are very specific to the company for which they were developed.

The indicators used to control a development project depend on the project's goals and the organizational environment. There is no default set of indicators that is always used in all development projects in the same manner. According to [14], a "good" indicator has to (a) support analysis of the intended information need, (b) support the type of analysis needed, (c) provide the appropriate level of detail, (d) indicate a possible management action, and (e) provide timely information for making decisions and taking action. The concrete indicators that are chosen should be derived in a systematic way from the project goals [12], making use of, for instance, the Goal Question Metric (GQM) approach [3]. Some examples from indicators used in practice can be found in [1]. With respect to controlling project cost, the Earned Value approach provides a set of commonly used indicators and interpretation rules. With respect to product quality, there exists even an ISO standard [10]. However, the concrete usage of the proposed measures depends upon the individual organization.

The test / diagnosis of complex systems was put on a formal basis in 1967 by [16]. One of the drawbacks was addressed by [13]. A good overview of system diagnosis models can be found in [2]. With respect to the visualization and applicable tools, an overview is presented in [20].

## 3   Goal-oriented Software Project Control

Specula [8] is a state-of-the-art SPCC. It interprets and visualizes collected measurement data in a goal-oriented way in order to effectively detect plan deviations. The control functionality provided by Specula depends on the underlying goals with respect to project control. If these goals are explicitly defined, the corresponding functionality is composed out of packaged, freely configurable control components. Specula provides four basic components: (1) a logical architecture for implementing software cockpits, (2) a conceptual model formally describing the interfaces between data collection, data interpretation, and data visualization [9], (3) an implementation of the conceptual model, including a construction kit of control components, and (4) a methodology of how to select control components according to explicitly stated goals and customize the SPCC functionality [8]. The methodology is based on the Quality Improvement Paradigm (QIP) and makes use of the GQM approach [3] for specifying measurement goals. QIP is used to implement a project control feedback cycle and make use of experiences and knowledge gathered in order to reuse and customize control components. GQM is used to drive the selection process of finding the right control components according to defined goals. The different phases that have to be considered for setting up and applying project control mechanisms can be characterized as follows:

*I. Characterize Control Environment:* First, stakeholders characterize the environment in which project control shall be applied in order to set up a measurement program that is able to provide a basis for satisfying all needs.

*II. Set Control Goals:* Then, measurement goals for project control are defined and metrics are derived determining what kind of data to collect. In general, any goal derivation process can be used for defining control objectives. For practical reasons, we focus on the GQM paradigm for defining concrete measurement goals addressing the measurement object, purpose, quality focus, viewpoint, and context information.

*III. Goal-oriented Composition:* Next, all control mechanisms for the project are composed based on the defined goals in order to provide online feedback on the basis of the data collected during project execution; that is, control techniques and visualization mechanisms are selected from a corresponding repository and instantiated in the context of the project that has to be controlled. This process is driven by interpretation and visualization models that clearly define which indicators contribute to specific control objectives, how to assess and aggregate indicator values, and how to visualize control objectives and intermediate results.

*IV. Execute Project Control Mechanisms:* Once all control mechanisms are specified, a set of role-oriented views is generated for controlling the project. When measurement data are collected, the control mechanisms interpret and visualize them accordingly, so that plan deviations and project risks are detected and a decision-maker can react accordingly. If a deviation is detected, its root cause must be determined and the control mechanisms have to be adapted accordingly. This, does, for example, require data analyses on different levels of abstraction in order to be able to trace causes of plan deviations.

*V. Analyze Results:* After project completion, the resulting visualization catena has to be analyzed with respect to plan deviations and project risks detected in time, too late, or not detected at all. The causes for plan deviations and risks that were detected too late or that were not detected at all have to be determined.

*VI. Package Results:* The analysis results of the control mechanisms that were applied may be used as a basis for defining and improving control mechanisms for future projects (e.g., selecting the right control techniques and data visualizations, choosing the right parameters for controlling the project).

Fig. 1 illustrates the basic conceptual modules of the Specula approach. The customization module is responsible for selecting and adapting the control components according to project goals and characteristics and defined measurement (control) goals. It is possible to include past experience (e.g., effort baselines, thresholds) in the selection and adaptation process. This experience is stored in a experience base. A Visualization Catena (VC) is created, which formally describes how to collect, interpret, and visualize measurement data. The set of reusable control components from which the VC is instantiated basically consists of integrated project control techniques (for interpreting the data in the right way) and data visualization mechanisms (for presenting the interpreted data in accordance with the role interested in the data). The central processing module collects measurement data during project performance and interprets and visualizes them according to the VC specification. Measurement data can be retrieved automatically from project repositories or manually from data collection forms and formal documents. Finally, charts and tables are produced to allow for online project control. A packaging module collects feedback from project stakeholders about the application of the control mechanisms and stores them in an Experience Base (e.g., whether a baseline worked, whether all plan deviations were detected, or whether retaliatory actions had a measurable effect). Using these modules, the Specula framework is able to specify a whole family of project control centers (which is comparable to a software product line for control centers).

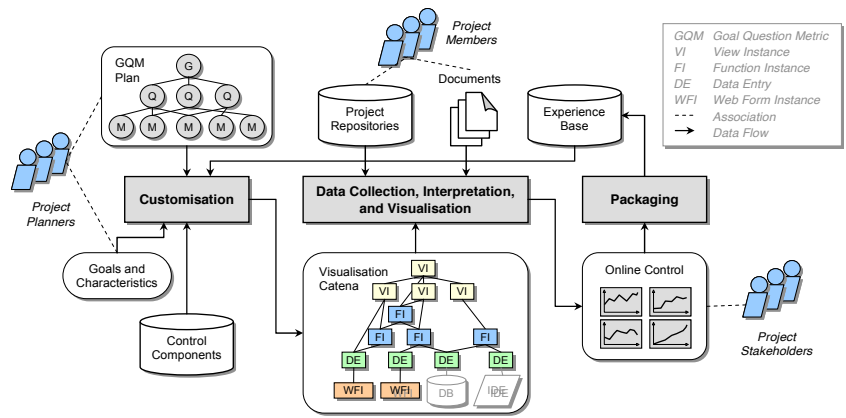

**Fig. 1.** Overview of the Specula framework.

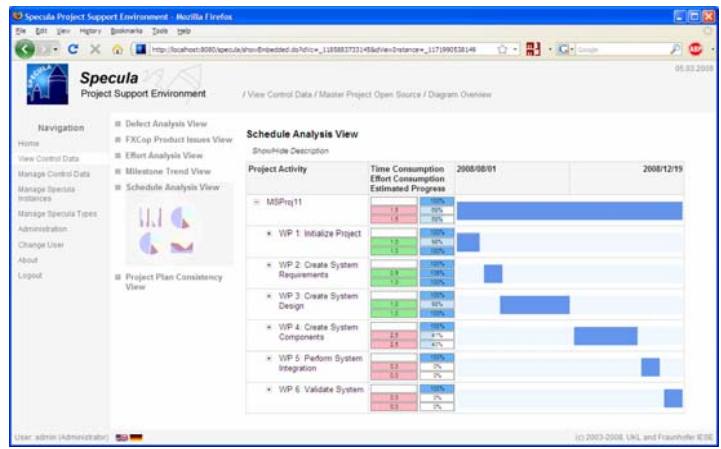

**Fig. 2.** Example visualization of a simple hierarchical Gantt chart.

The Specula approach was evaluated as part of industrial case studies in the Soft-Pit project (a public German research project, no. 01ISE07A) in which a prototypical implementation of the concepts was used. Results of the first two iterations can be found in [5] and [6]. In general, people perceived the usefulness and ease of use of the Specula control center as positive. However, usefulness and ease of use also varied across the different case study providers depending on the state of the practice before introducing the control center solution and it also largely varied across the different visualization mechanisms used. In the Soft-Pit case, mostly "standard visualizations" for project control were used, such as Gantt charts, line/bar charts, tables/matrixes and simple trees (see, e.g., Fig. 2). One major success factor for the usefulness of visualizations was how intuitive the visualization can be interpreted. Especially, when aggregating data and complex, multi-dimensional relationships needs to be illustrated, this requires more advanced visualization concepts.

## 4      Advanced Visualization Mechanisms

The following sub-sections present examples for advanced mechanisms used for visualizing the risks and quality of development projects and summarize lessons learned from their application. Further visualization mechanisms of quality properties, especially safety and security for embedded systems, is currently investigated in the German research project ViERforES (see *http://www.vierfores.de*).

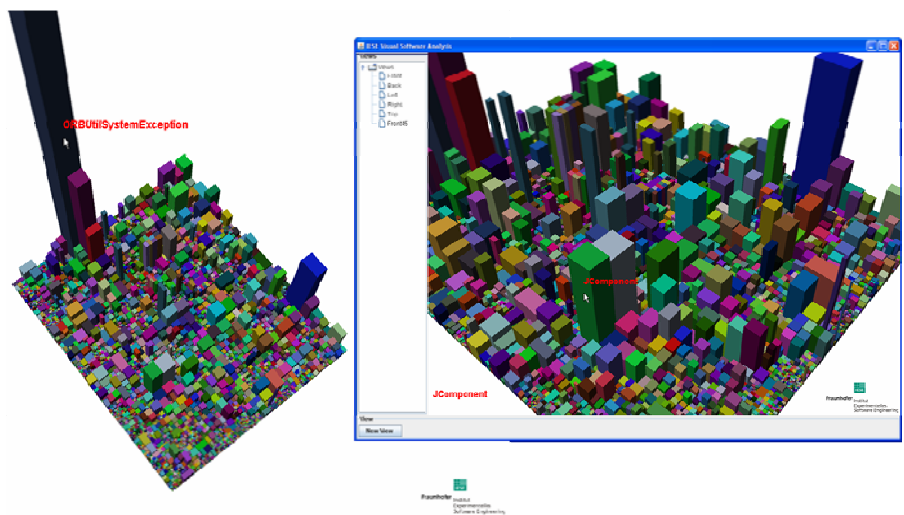

**Fig. 3.** 3D Treemap visualizing different metrics.

### 4.1      Visualizing Code Quality

To analyze the quality of a software system, metrics are used that measure certain attributes of the software's internal structure. Many metric tools exist that are able to define such metrics and collect measurement data automatically. For analysis purposes different visualization techniques such as node-link diagrams and graphs are used in order to help the user in drawing conclusions about the quality of the software system. These techniques use a limited set of graphical elements like text, simple geometric shapes or uniform color fills to highlight relevant attributes of the software system being visualized. For combining different metric values within one picture, a 3D-Treemap technique (see Fig. 3) was developed and integrated into a code analysis system at Fraunhofer IESE. This visualization mechanism allows us to map data measuring code quality to different graphical properties of each cube (such as position, size, height, and color). To further analyze these values, the user is able to interactively define new views, pan and zoom within the 3D scene and use a pull-down menu to initialize other measurement or visualization actions.

### 4.2  Visualizing Risk Management

Risk management and especially risk avoidance plays an important role in all development and construction activities. For managing risks, a structured process is mandatory. Visualization is used for analyzing risks and supporting managers in deciding upon necessary actions. Siemens developed a methodology named sira that is used for collecting data about possible risks [4]. This includes structured interviews for determining possible risks, their probability and importance as well as the possible damage that may be caused. Based on this analysis, a risk portfolio is created. In order to analyze these risks, the so-called sira bubble charts were created that summarize all necessary information that has to be discussed with the customer (see Fig. 4 and [4]).

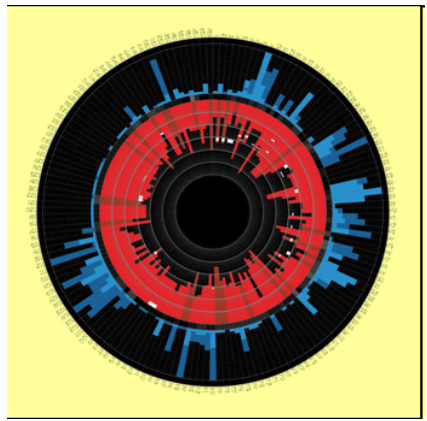

**Fig. 4.** sira.iris, a visualization of a risk portfolio.

### 4.3  Fault Detection in Distributed Systems

The functionality of a software system is distributed over many components and the interaction between these components plays a crucial role. In order to analyze the reliability of a system, data about communicating components and their error-proneness is collected. The output is analyzed using an interactive visualization (see Fig. 5 and [20]). In this visualization color coding is used to characterize the "faultiness" of components with respect to communication relations. The ratio between faulty communication and the overall amount of communication is used for coloring all nodes and edges of the graph, indicating starting points for bug fixing activities. The overall approach is described in [11]. Interaction plays an important role in this application. For instance, changing colors helps understanding the impact of faults. Changing transparency of clusters helps understanding structural information about the system.

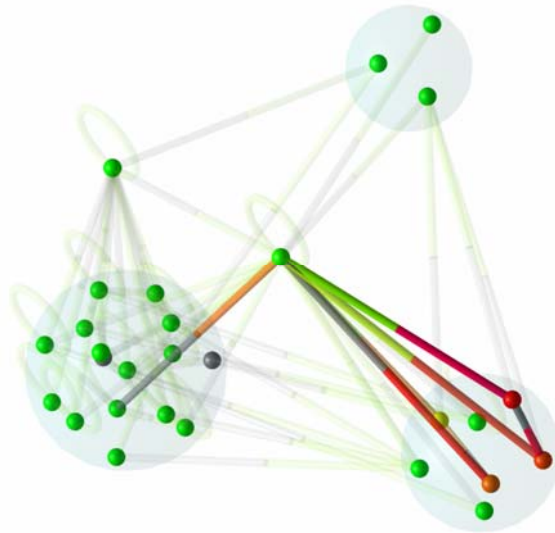

**Fig. 5.** Faults collected during the execution of a system.

## 5    Conclusions

This article presented the basic concept of an SPCC for establishing project control by means of systematic visualization mechanisms. We illustrated existing approaches and presented a goal-oriented way to establish project control by formalizing the way measurement data are interpreted and visualized according to a previously defined measurement goal. Existing approaches offer mostly partial solutions. Especially goal-oriented usages based on a flexible set of techniques and methods are not comprehensively supported [15]. The expected benefits of the goal-oriented visualization approaches include: (1) improvement of quality assurance and project control by providing a set of custom-made views of measurement data, (2) support of project management through early detection of plan deviations and proactive intervention, (3) support of distributed software development by establishing a single point of control, (4) enhanced understanding of software processes, and improvement of these processes, via measurement-based feedback, and (5) preventing information overload through custom-made views with different levels of abstraction.

An important research issue in this context is the development of a schema for adaptable control techniques and methods, which effectively allows for purpose-driven usage of an SPCC in varying application contexts. Another research issue is the elicitation of information needs for the roles involved and the development of mechanisms for generating adequate role-oriented visualizations of the project data. Another important research issue is support of change management. When the goals or characteristics of a project change, the real processes react accordingly. Consequently, the control mechanisms, which should always reflect the real world situation, must be updated. This requires flexible mechanisms that allow for reacting to process variations. One long-term goal of engineering-style software development is to control and forecast the impact of process changes and adjustments on the quality of the software artifacts produced and on other important project goals. Goal-oriented visualization mechanisms can be seen as a valuable contribution towards reaching this goal.